# Absolute Doubly Differential Angular Sputtering Yields for 20 keV $Kr^+$ on Polycrystalline Cu

Caixia Bu[1*], Liam S. Morrissey[2,3*], Benjamin C. Bostick[4], Matthew H. Burger[5], Kyle P. Bowen[1], Steven N. Chillrud[4], Deborah L. Domingue[6], Catherine A. Dukes[7], Denton S. Ebel[8], George E. Harlow[8], Pierre-Michel Hillenbrand[1], Dmitry A. Ivanov[1], Rosemary M. Killen[2], James M. Ross[4], Daniel Schury[1], Orenthal J. Tucker[2], Xavier Urbain[9], Ruitian Zhang[1], and Daniel W. Savin[1*].

[1]Columbia Astrophysics Laboratory, Columbia University, New York, NY 10027, USA

[2]NASA Goddard Space Flight Center, Greenbelt, MD 20771, USA

[3]Faculty of Engineering and Applied Science, Memorial University, NL, Canada A1B 3X7

[4]Lamont-Doherty Earth Observatory, Columbia University, Palisades, NY 10964, USA

[5]Space Telescope Science Institute, 3700 San Martin Dr., Baltimore, MD 21218, USA

[6]Planetary Science Institute, Tucson, AZ 85719, USA

[7]Laboratory for Astrophysics and Surface Physics, University of Virginia, Charlottesville, VA 22904, USA

[8]American Museum of Natural History, New York, NY 10024, USA

[9]Université catholique de Louvain, B-1348 Louvain-la-Neuve, Belgium

*Corresponding Authors: cb3619@columbia.edu; lsm088@mun.ca; dws26@columbia.edu.

**Abstract**

We have measured the absolute doubly differential angular sputtering yield for 20 keV $Kr^+$ impacting a polycrystalline Cu slab at an incidence angle of $\theta_i = 45°$ relative to the surface normal. Sputtered Cu atoms were captured using collectors mounted on a half dome above the sample, and the sputtering distribution was measured as a function of the sputtering polar, $\theta_s$, and azimuthal, $\phi_s$, angles. Absolute results of the sputtering yield were determined from the mass gain of each collector, the ion dose, and the solid angle subtended, after irradiation to a total fluence of $\sim 1 \times 10^{18}$ ions/cm$^2$. Our approach overcomes shortcomings of commonly used methods that only provide relative yields as a function of $\theta_s$ in the incidence plane (defined by the ion velocity and the surface normal). Our experimental results display an azimuthal variation that increases with increasing $\theta_s$ and is clearly discrepant with simulations using binary collision theory. We attribute the observed azimuthal anisotropy to ion-induced formation of micro- and nano-scale surface features that suppress the sputtering yield through shadowing and redeposition effects, neither of which are accounted for in the simulations. Our experimental results demonstrate the importance of doubly differential angular sputtering studies to probe ion sputtering processes at a fundamental level and to explore the effect of ion-beam-generated surface roughness.

# 1 Introduction

Sputtering of atoms from surfaces by ion irradiation is important for a diverse range of fields, such as thin film production for optical coatings and electronic devices,[1–4] the design of containment vessels for fusion reactors,[5–7] surface processing and analysis,[8–11] nanofabrication,[12–16] and studies of surface and exosphere evolution of airless planetary bodies such as Mercury, the Moon, and asteroids.[17–19] Our ability to advance these fields requires knowledge of the absolute polar and azimuthal distributions of the sputtered atoms. However, our understanding of sputtering at this fundamental level remains incomplete. Although the polar distribution of sputtered atoms has been well studied experimentally and theoretically, that is not the case for the azimuthal distribution, which is extremely challenging to study experimentally.[20,21]

The ideal experimental approach for measuring doubly differential angular sputtering yields (defined as the number of sputtered atoms per incident ion per steradian [sr]) is to use collectors distributed over a hemisphere to capture the sputtered atoms. Cylindrical and planar collectors suffer from insolvable geometric distortions.[20,22] One challenge with using the collector methodology, though, has been to reliably convert the deposited film thickness into a number of atoms. For example, absolute sputtering results for single crystal metal targets have been reported using electron backscattering to determine the deposition thickness.[22–24] However, converting this to a total atom number requires knowledge of the exact atomic arrangement and corresponding density of the film deposited.[25] This information is not reported in Refs. [22-24], resulting in an undefined uncertainty in their results. Relative measurements for sputtering from polycrystalline metal targets have also been reported using a hemispherical collector.[26] These show good qualitative agreement with state-of-the-art theory,[21] but are of limited utility due to their relative nature. Another absolute approach is to use a moveable quartz crystal microbalance (QCM) as an in-situ collector. However, the only such work of which we are aware[27] needed to rotate the plane of the sample in order to vary the azimuthal collection angle. That approach is not suitable for loose powder samples.

Here, we report absolute measurements for the doubly differential angular sputtering yields for 20 keV $Kr^+$ impacting a polycrystalline Cu substrate at an angle of incidence $\theta_i = 45°$. We use a half-dome collection geometry and provide absolute yields by determining the number of sputtered atoms from mass-gain measurements. This avoids the issues of geometric distortion, unknown deposited material density, and will enable us to eventually study loose powders. Below, we discuss the apparatus and experimental methods (Section 2), the theoretical calculations (Section 3), the experimental and theoretical results (Section 4), and the implications of our findings (Section 5).

## 2 Experimental Methodology

### 2.1 Apparatus Description

Measurements were performed using an ion beam apparatus capable of irradiating both loose powder and solid targets, necessitating irradiation of the samples from above.[28] The ion source and first leg of the ion beam line stand ~ 2 m above the sample stage, the latter of which is located within the target chamber. Ions were produced in a duoplasmatron source,[29] extracted electrostatically, and accelerated to an energy of 20 keV. The desired $Kr^+$ was charge-to-mass selected using a Wien filter. The resulting $Kr^+$ beam was transported electrostatically to the end of the first leg of the apparatus and into a 90° spherical electrostatic deflector that directed the ions onto a trajectory with a polar impact angle of $\theta_i = 45°$ relative to the normal of the sample surface. The definition of the angles used here are shown in Fig. 1. After this bend, at the beginning of the second leg of the apparatus, the ion beam was shaped and steered using an einzel lens and a set of XY deflectors. The apparatus up to this point is nearly identical to that described by Bruhns et al.,[30] which was modified into the configuration shown in Bu et al.[28]

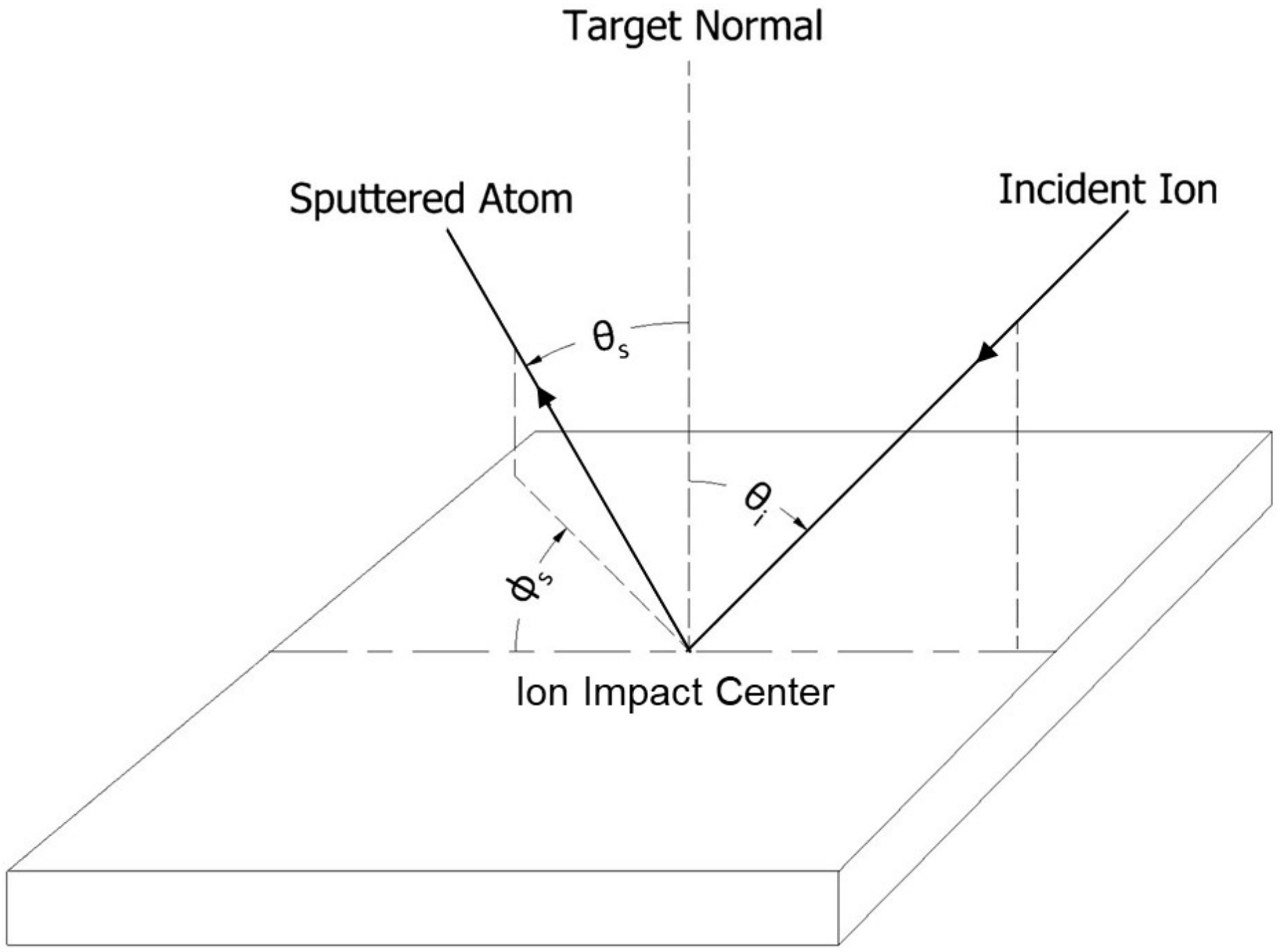

**Fig. 1** Definition of the angles. The ions impact the target at an incidence angle of $\theta_i$ = 45° with respect to the target normal. Atoms are sputtered at a polar angle $\theta_s$ relative to the target normal. The azimuthal angle $\phi_s$ of the sputtered atoms is measured clockwise with respect to ion beam direction projected onto the target.

In order to reproducibly deliver the ion beam onto the target, the ions then passed through two 5 mm collimating apertures spaced 1.2 m apart. These geometrically constrained the shape and position of the ion beam on the sample. The resulting divergence of the impact angle on the target was ±0.23° in the incidence plane (the plane containing the ion velocity vector and the normal to the sample surface) and ±0.16° in the perpendicular plane passing through the sample center. The maxima of the beam spot on the target were 15.1 and 10.7 mm, respectively, due to the geometric constraints from the collimating apertures.

After the second aperture and before the gate valve to the target chamber, we used a single-wire beam profile monitor (BPM)[31] to monitor the ion beam current, shape, and position. The BPM current measurement was calibrated using a Faraday cup positioned behind the sample location. The beam transmittance from the BPM to this cup was 100%, enabling us to calibrate the BPM with a reproducibility of 20%. We attribute this systematic uncertainty to changes in the surface properties of the BPM wire due to ion irradiation over the course of the measurement campaign. Here and throughout, all uncertainties are quoted at a one-sigma confidence level.

In the target chamber, the position of the ion beam on the sample stage was determined using an alumina scintillator, the image of which was recorded in 60-s exposures with a digital camera. The center of the ion beam was within 1 mm of the center of the sample stage. The scintillator image also enabled us to calibrate the beam position as recorded with the BPM to that on the sample stage. Projecting the BPM measurements of the beam shape onto the sample stage, and accounting for $\theta_i$, the ion beam full width at half maximum (FWHM) on the target was ~ 10 mm in the incidence plane and ~ 7 mm in the perpendicular plane.

With the ion beam blocked by the gate valve and before beginning of each day's irradiation, we used the BPM to verify that the beam was properly positioned and that the ion current was sufficiently high to deliver about one-tenth of the desired total dose of the campaign. Typical currents were ~ 0.2 – 0.4 μA, corresponding to fluxes of ~ (2.3 – 4.6) × $10^{12}$ ions/s/$cm^2$, using the ion beam FWHM on the target. During irradiation, the BPM readings were recorded every second.

To compensate for a typical slow decrease in the ion current with time, moderate retuning of the beam was sometimes required over the course of a day, during which the beam was blocked. The sample was irradiated for a total of $3.11 \times 10^5$ s, over a period of 10 non-consecutive days, resulting in a total delivered dose of $(6.20 \pm 1.24) \times 10^{17}$ ions, for an overall fluence of $\sim 1 \times 10^{18}$ ions/cm$^2$. The systematic uncertainty in the dose is due entirely to the BPM calibration. The total dose delivered was selected so that the mass gain on the majority of the collectors was well above the accuracy of the weighing system (see below). During irradiation, the pressure in the target chamber was $\sim 7 \times 10^{-9}$ Torr. The base pressure with the gate valve closed was $\sim 5 \times 10^{-10}$ Torr.

The target used was polycrystalline Cu of 99.99% purity, which was hand polished and ultrasonically cleaned in an ethanol bath. The Cu sample was 1.6 mm thick and cut into a 50.8 mm × 50.8 mm slab. The arithmetical mean surface roughness of the slab was $Ra \sim 0.1$ µm, as characterized using three-dimensional digital light microscopy.[32]

## 2.2 Mass-Gain Measurements

Sputtered atoms were collected using circular gold-coated quartz crystals. The collectors used are standard components from INFICON Inc. (Bad Ragaz, Switzerland; model: 008-010-G10). Each collector has a diameter of 14 mm and is coated with diffuse gold (Au) on one side, which provides a high, stable sticking coefficient over a wide range of temperatures (see Sec. 3.1). The average mass of the collectors without any deposition was 89.50 ± 0.50 mg. The collectors were mounted on a half dome centered above the sample stage (see Fig. 2), so that the Au-coated side faced the Cu target. The radial distance from the sample center to each collector center was 70 mm. The circular area of each collector exposed to the target was 10 mm in diameter, spanning a solid angle of $1.60 \times 10^{-2}$ sr. One collector was situated at a polar sputtering angle of $\theta_s = 0°$, spanning all azimuthal sputtering angles $\phi_s$. This was surrounded by rings of collectors at $\theta_s = 15°$, 30°, 45°, 60°, and 75° (Fig. 2(a)). The centers of the collectors in the ring at $\theta_s = 15°$ spanned $\phi_s$ from 0° to 180°; at $\theta_s = 30°$, $\phi_s$ spanned from 0° to 142°; at $\theta_s = 45°$, $\phi_s$ spanned from 0° to 143.4°; at $\theta_s = 60°$, $\phi_s$ spanned from 0° to 153°; and at $\theta_s = 75°$ $\phi_s$ spanned from 0° to 159.5° (Fig. 2(b), see also Table S1 in the Supplementary Material). Here, $\phi_s = 0°$ is the direction of the ion beam in the incidence plane.

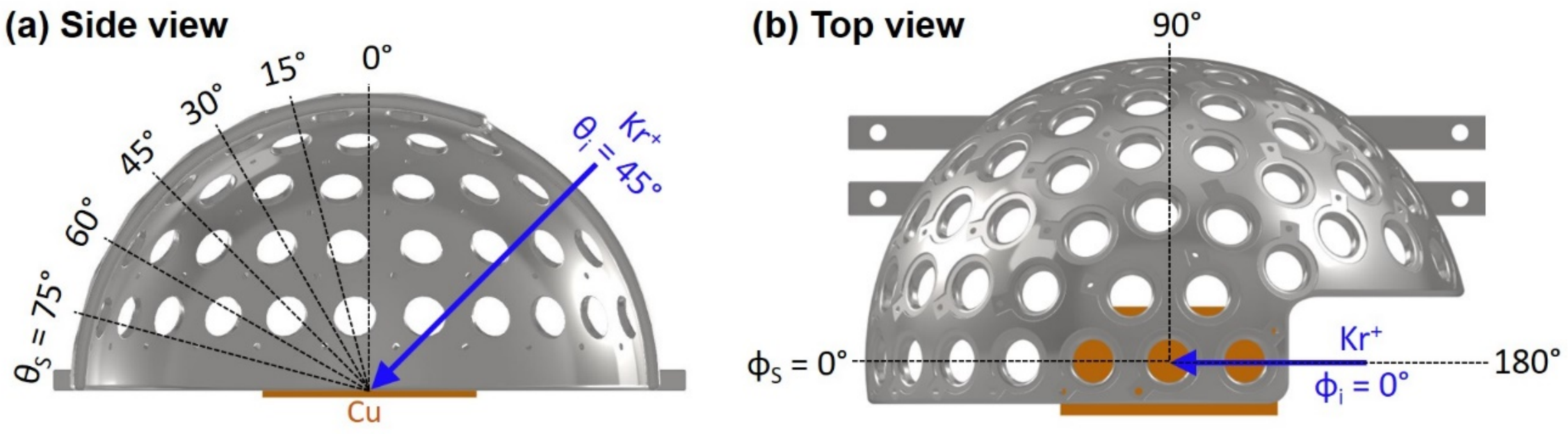

**Fig. 2.** Schematic of the half-dome mount for the collectors and the Cu target. The collectors are installed in the openings shown in the half dome. Each opening has a diameter of 10 mm. The radial distance from the sample center to each collector center is 70 mm. The Cu target is mounted horizontally. (a) Side view showing the polar angle $\theta_i$ of the incident $Kr^+$ beam (blue arrow) and the polar angles $\theta_s$ of the collectors. (b) Top view showing the azimuthal angle $\phi_i$ of the $Kr^+$ beam velocity vector (blue arrow) and the azimuthal angles $\phi_s$ for the collectors.

Absolute doubly differential sputtering yields were determined from the mass gain of each collector (converted to the number of Cu atoms) divided by the total ion dose delivered and by the solid angle subtended. The mass of each collector was measured individually ex-situ before and after irradiation using a QCM (Mettler Toledo, LLC, Columbus, OH, USA; model: UMX2 Ultra-microbalance) that was integrated into an automated weighing system in a HEPA (high efficiency particulate air) filtered environmentally controlled chamber (Measurements Technology Limited, Minneapolis, MN, USA). Fig. 3 shows images of a collector before and after ion irradiation of the Cu target. The collectors were left in the environmentally controlled chamber to equilibrate for 24 hours prior to weighing and six Po-210 strips were arranged around the weighing cage for electrostatic control.

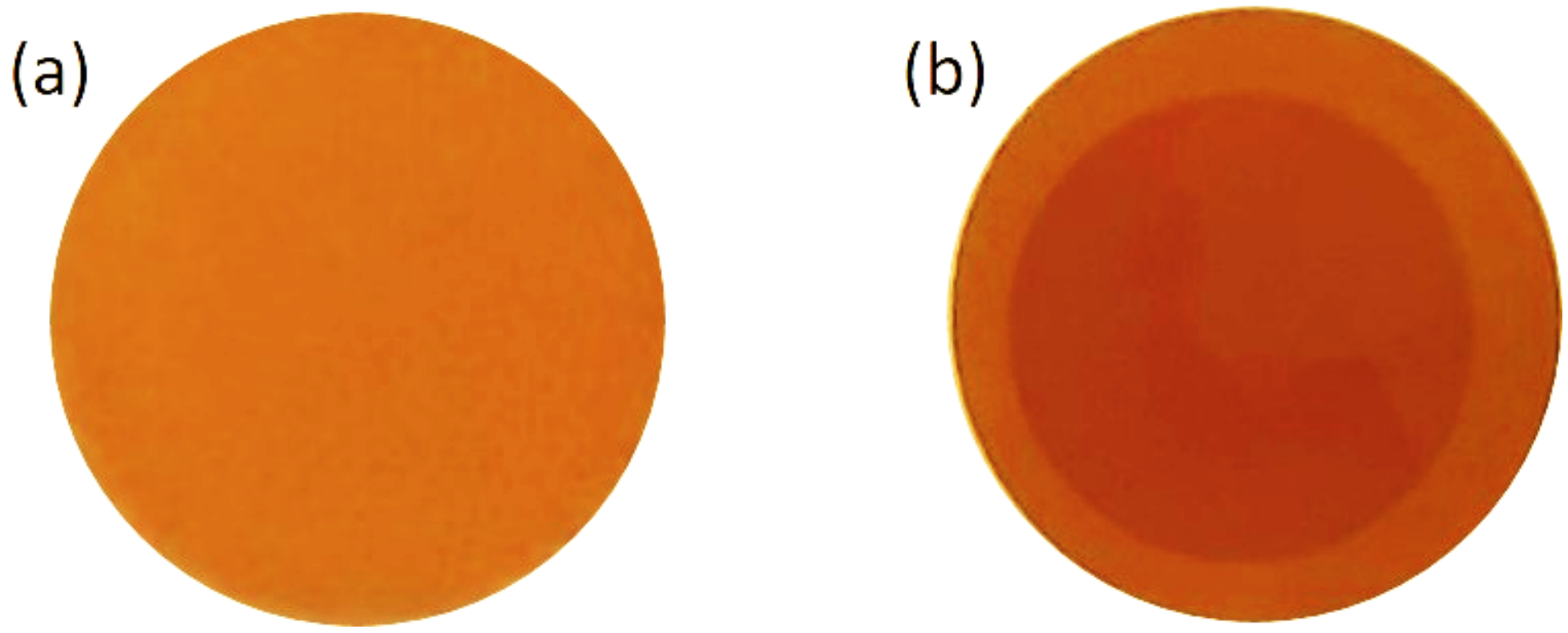

**Fig. 3** Images of the 14 mm diameter collector located at $\theta_s = 0°$, taken by a digital camera (a) before and (b) after the irradiation. The contrast in the images has been artificially enhanced to better display the Cu-coated region. The mass of this collector before irradiation was 89.97204 ± 0.00012 mg, and the mass gain after irradiation was 10.12 ± 0.13 µg. The darker area in (b) has a diameter of 10 mm and is due to the deposition of sputtered Cu atoms onto the Au surface.

The mass of each collector was measured in triplicate at three separate points in time, before and after irradiation. Buoyancy corrections were applied for any changes in barometric pressure at the times of weighing. Averages of the before and after collector masses were used to derive the mass gains for sputtering yield measurement and corresponding uncertainties. The total time between the initial weighing of the collectors and the final weighing was 35 days. In addition, four collectors with no direct line of sight to the target were mounted in the target chamber. These witness collectors were used to monitor for any mass change from the handling of the collectors before, during, and after the irradiation. No such changes were observed.

# 3 Theoretical Modelling

## 3.1 Binary Collision Approximation

Monte Carlo simulations using the binary collision approximation (BCA) were conducted for comparison to the experimental results. BCA modelling was chosen as it has been shown to have good agreement with experimental results for total sputtering yields[21,33] and for the polar angle distribution of the sputtered atoms in the incidence plane[34] at the impact energy studied here.

The BCA approach treats sputtering as a result of binary collision cascades involving the projectile and target atomic nuclei. Simulations were conducted using SDTrimSP (Version 6.06),[35] an extension of the Transport of Ions in Matter (TRIM) software, which can be run in standard (S) or dynamic (D) modes using serial (S) or parallel (P) processing. SDTrimSP is a state-of-the-art BCA code, tracking the kinetic energy of the sputtered and substrate atoms. This version does not account for the effects of atomic ionization, ionic neutralization, potential sputtering, substrate crystallinity, or surface roughness. Molecular dynamics (MD) calculations are often used to address some of these shortcomings. However, the required simulation substrate size and time step increase with impact energy, making MD calculations computationally prohibitive at energies above a few keV.[36]

SDTrimSP simulations were conducted for a ray of 20 keV Kr (SDTrimSP can only simulate neutral impactors) impacting a flat amorphous Cu surface at $\theta_i$ = 45°. A total of $10^5$ impacts were simulated onto a 1000 Å thick Cu slab. No significant difference was observed between the static and dynamic simulations, the latter of which accounts for implantation of the Kr. For the surface binding energy of Cu, we used its monatomic cohesive energy, 3.49 eV, which has been shown to accurately predict the total sputtering yield and energy distribution of Cu as compared to experiment for impact energies above 1 keV.[36] The energies and the polar and azimuthal emission angles of the sputtered atoms were recorded for each Kr impact. Each sputtered atom was then mapped onto the half dome and recorded if it intersected a collector. This enabled us to simulate the sputtering yield onto each collector.

## 3.2 Molecular Dynamics

In order to derive accurate sputtering yields for each collector, we need to know the sticking coefficients for Cu on Au and Cu surfaces. The Au represents a fresh collector and the Cu a collector that has developed at least one monolayer (ML) of Cu. Here, we have calculated the corresponding sticking coefficients with molecular dynamics (MD) simulations using the Large-scale Atomic/Molecular Massively Parallel Simulation (LAMMPS) package.[37] Interactions between all atoms were simulated using an embedded atom method (EAM) interatomic potential designed specifically for the Cu-Cu and Cu-Au systems.[38] For each case we simulated the sputtered Cu atoms using a Thompson[39] energy distribution, shown in Fig. 4(a), for binned energies from 0.1 – 24.0 eV, shown in Fig. 4(b). The surface binding energy of the polycrystalline Cu from which the Cu atoms were sputtered was taken to be the Cu monatomic cohesive energy. We simulated impacts at polar angles of both 0° and 45° relative to the surface normal, the latter to simulate surface roughness,[40,41] and with a random azimuthal angle. The Cu and Au targets were modelled as single crystals at room temperature and oriented, using Miller indices, with the [100], [010], and [001] directions along the x, y, and z axis, respectively. One hundred Cu atoms were simulated at each energy for statistics. Cu was considered 'stuck' to the surface if after 2 ps (10000 time steps) its final position was within the length of one Cu-surface bond, 2.7 and 2.5 Å for Cu-Au and Cu-Cu bonding, respectively. Those atoms that did not stick reflected off the surface back into the vacuum region. As shown in Fig. 4(b), the sticking coefficient of Cu onto Au was dependent on the Cu emission energy from the $Kr^+$-irradiated sample. The energy-distribution-averaged Cu-Au sticking coefficient was 82% and 74% for 0° and 45° impact angles, respectively. Thus, we take (78 ± 4)% for the energy-distribution-weighted average Cu-Au sticking coefficient. This contrasts with the Cu-Cu sticking coefficient, which was 100% across the entire ejecta energy distribution. Sticking coefficients were also calculated at 350 K. This is the maximum expected sample temperature, based on the power delivered by the ions and the thermal capacity of the Cu sample. Taking into account radiative and conductive cooling of the target would result in a lower maximum sample temperature. The temperature of the collectors, which are 7 cm away from the target, are expected to be lower than the sample temperature. The 350 K sticking coefficients were identical to the values at room temperature. The measured sputtering yields were calculated assuming a 100% Cu-Au sticking coefficient. In reality, the 78% Cu-Au sticking coefficient means that an additional (1/0.78-1) ML of Cu atoms must be sputtered before a full ML of Cu builds up on the Au. Using the average planar density for the (100), (110), and (111) crystallographic planes for face centered cubic Cu, we approximated one Cu ML as $1.5 \times 10^{15}$ atoms/cm$^2$. Thus, the assumed 100% sticking coefficient for the first ML results in an underestimation of 0.28 ML × $1.5 \times 10^{15}$ atoms/cm$^2$ × 0.785 cm$^2$ = $0.332 \times 10^{15}$ sputtered atoms onto each collector, given that the radius of the exposed area of each collector is 0.5 cm. Dividing the number of sputtered atoms by the total dose of $(6.20 \pm 1.24) \times 10^{17}$ ions and the solid angle of $1.60 \times 10^{-2}$ sr of each collector, this corresponds to an underestimation in the sputtering yield onto any individual collector of 0.033 atoms/ion/sr. As we show below, this is insignificant for all but a few of the collectors.

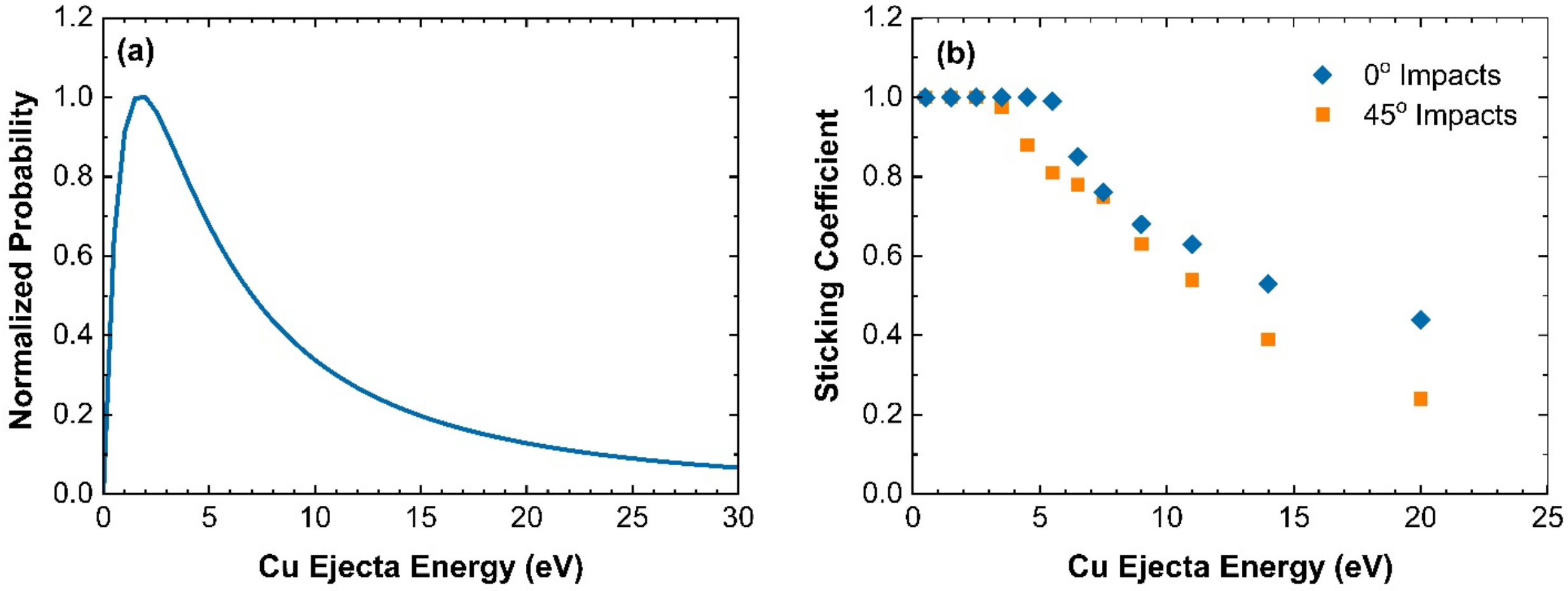

**Fig. 4** (a) Normalized sputtered Cu energy distribution using the Thompson distribution; (b) Cu sticking coefficient onto Au as a function of Cu ejecta energy.

## 4 Results and Discussion

### 4.1 Absolute Doubly Differential Angular Sputtering Yields

The absolute doubly differential angular sputtering yields from our experimental measurements are presented in Fig. 5(a) and Table S1. The measured mass gains of the collectors were up to 10 μg. The average uncertainty in the mass gain was 0.143 ± 0.141 μg, corresponding to a sputtering yield uncertainty of 0.136 ± 0.134 atoms/ion/sr. As discussed above, we have assumed a 100% sticking coefficient for the sputtered Cu onto the collectors, which introduces an underestimate in the sputtering yield for each collector of 0.033 atoms/ion/sr. The uncertainties in the mass gain measurement and the sticking coefficient introduce an insignificant uncertainty in our results for all but a few collectors at $\theta_s = 75°$.

The SDTrimSP simulations are shown in Fig. 5(b) and Table S1. To estimate the effect of the finite beam size on the simulation, the position of the incoming Kr ray was shifted to the extreme edges of the beam spot in both the incidence plane and the perpendicular plane. Averaging the results for these four cases produced sputtering yields on each collector that were within ± 4% of the central impact calculation.

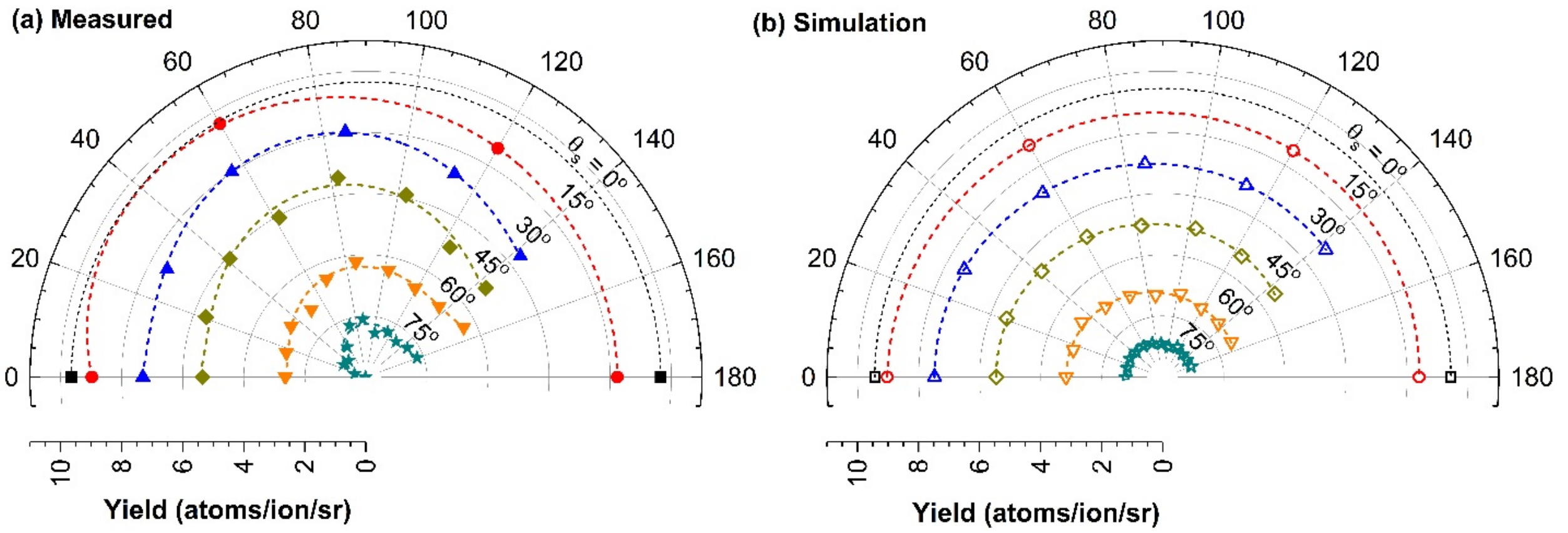

**Fig. 5** Absolute doubly differential angular sputtering yields, derived from our (a) measurements and (b) SDTrimSP simulations, as a function of the sputtering azimuthal angle $\phi_s$, with $\phi_s = 0°$ defined as the projection of the $Kr^+$ velocity vector onto the sample. The number next to each data set gives the corresponding sputtering polar angle $\theta_s$, with respect to the sample normal. For $\theta_s = 0°$, there was only one collector, which spanned all $\phi_s$. For the other values of $\theta_s$, each data point

represents one collector. The dashed lines are polynomial fits to the data. The fewer experimental points for large $\phi_s$ is due to the opening in the half dome for the ion beam to pass through (see Fig. 2). See also Table S1.

The measured results display several noteworthy features, the most obvious of which is the pronounced azimuthal anisotropy outside the incidence plane, as compared to the simulated results. As shown in Fig. 6, the deviation between the measured and simulated values grows with increasing $\theta_s$ and is most pronounced for azimuthal angles $\phi_s$ outside of the incidence plane. This result would not have been detected with standard experimental methods using cylindrical and planar collectors for differential sputtering yields, which are only able to reliably collect data in the incident plane (i.e., for $\phi_s$ along 0° and 180°).[20,42–44]

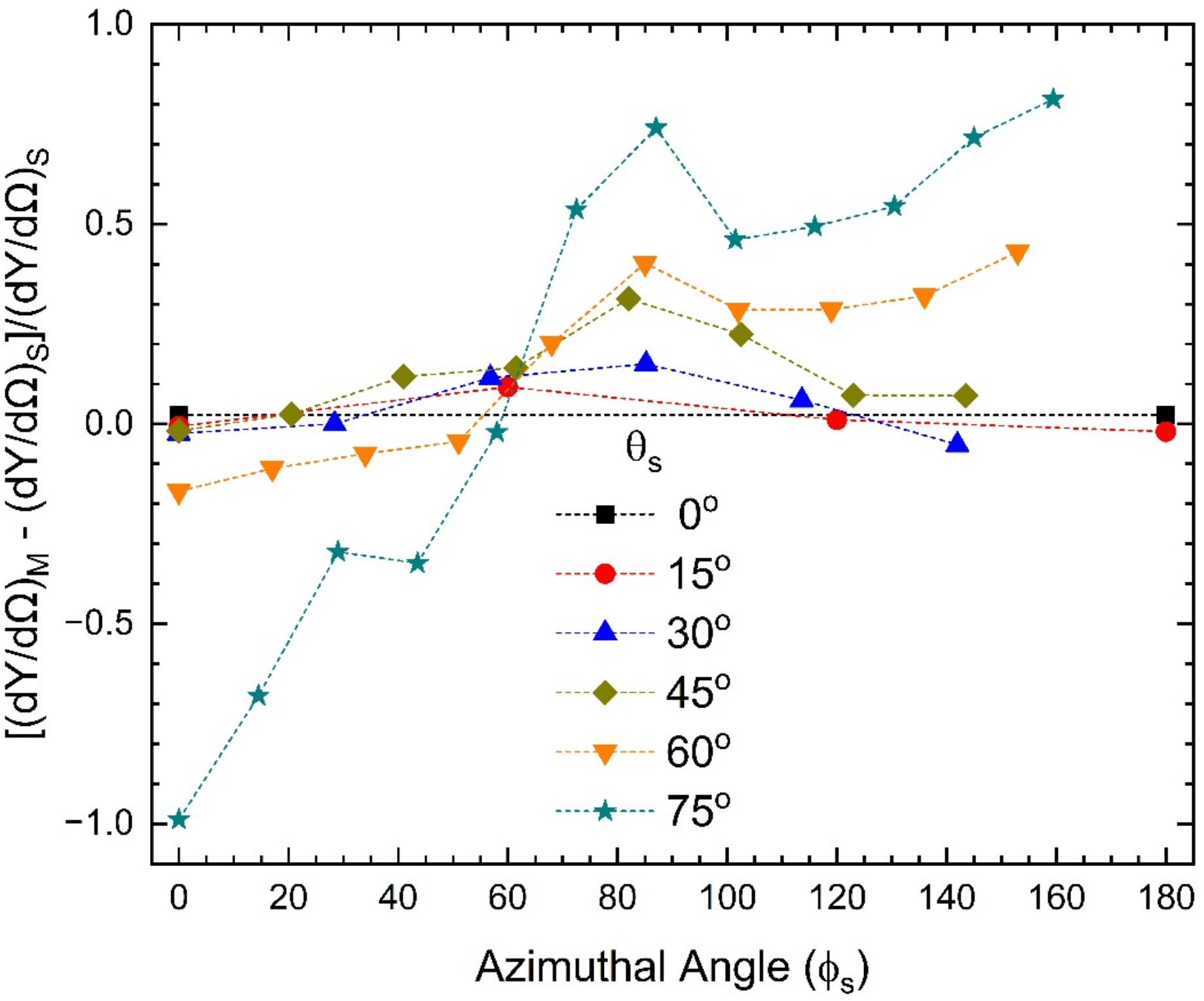

**Fig. 6** Difference between the measured $(dY/d\Omega)_M$ and simulated $(dY/d\Omega)_S$ doubly differential sputtering yields, as a function of sputtering azimuthal angle $\phi_s$ and scaled to $(dY/d\Omega)_S$. Each set of symbols is for one sputtering polar angle $\theta_s$. The dashed lines are to guide the eye. The $(dY/d\Omega)_M$ and $(dY/d\Omega)_S$ values are from Fig. 5(a) and Fig. 5(b), respectively. See also Table S1.

Considering first the expected results in the incidence plane, many previous simulations and experimental studies have shown an anisotropic distribution of the ejecta in the incidence plane, exhibiting enhanced sputtering in the forward direction ($\phi_s = 0°$) as compared to the backward ($\phi_s = 180°$).[43,45–48] However, those results were for impactors with comparatively low impact energies, where a large proportion of the sputtered atoms are expected to result from primary knock-on collisions and thus be effected by the momentum of the oblique impactor. For our experimental results in the incidence plane, as shown in the Fig. 7, the sputtering lobe peaks nearly normal to the surface with a tilted angle of $\theta_s \approx 2°$. This is similar to previous experimental studies for Si sputtered from a polycrystalline Si target by 20 keV $Kr^+$ at $\theta_i = 45°$.[46] For both experiments, there were a sufficient number of collisions in the substrate to create randomized secondary knock-on collisions. Sputtered atoms produced by these secondary knock-ons lose "memory" of the momentum of the impactor. However, as shown in Fig. 7, the collision cascade is not fully randomized yet, and there is a slight forward-backward anisotropy that can be seen in both experimental measurements and SDTrimSP simulations.

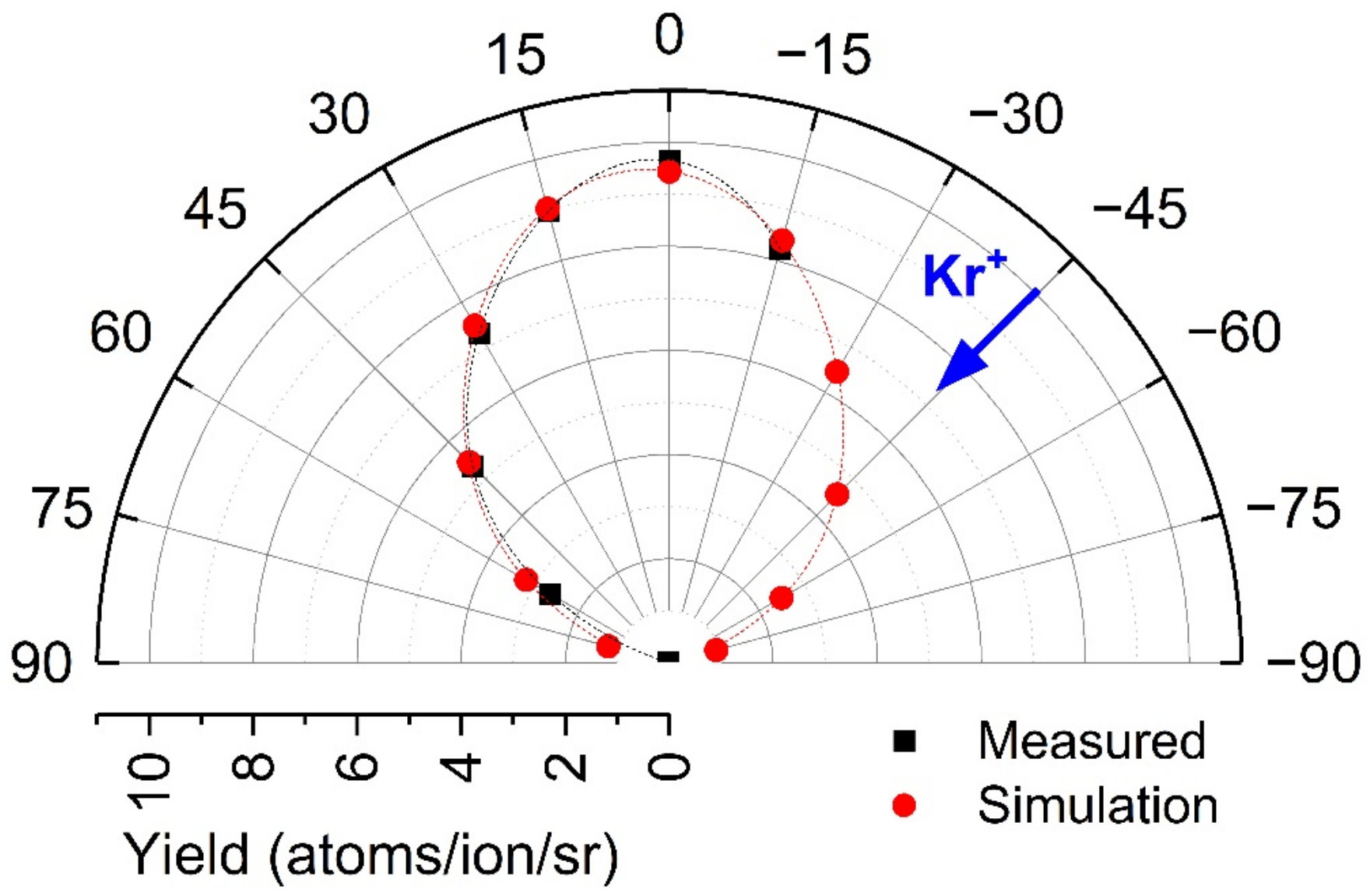

**Fig. 7** The measured $(dY/d\Omega)_M$ and simulated $(dY/d\Omega)_S$ doubly differential sputtering yields in the incidence plane as a function of sputtering polar angle ($\theta_s$). Positive $\theta_s$ indicates the forward direction ($\phi_s = 0°$), and the negative $\theta_s$ indicates the backward direction ($\phi_s = 180°$). The $(dY/d\Omega)_M$ and $(dY/d\Omega)_S$ values are from Fig. 5(a) and Fig. 5(b), respectively. See also Table S1. The dashed lines represent a polynomial interpolation between the data points.

The measured azimuthal distribution is moderately anisotropic for polar angles of $\theta_s = 15°$, 30°, and 45° and becomes significantly anisotropic for 60° and 75°. The anisotropy also appears as an apparent migration in the azimuthal peak for the sputtering yield from $\phi_s \sim 50°$ for $\theta_s = 15°$ to $\phi_s \sim 90°$ for $\theta_s = 75°$. Given the uniformity of the ion beam as measured by the BPM (the doses delivered to each quadrant of the beam spot were equivalent to within 3%), none of these features can be attributed to the ion beam profile. Similar peak migration has been seen previously for sputtering due to primary knock-on collisions,[26] but that cannot explain our results as the sputtered atoms here are due to secondary knock-on events. Taken all together, these findings point to the influence of surface structure on the azimuthal distribution of our experimental results. The formation of surface structure has also been noted in previous experimental results for fluences above ~ $10^{17}$ ions/cm$^2$.[13,49] Our fluence here is an order of magnitude greater.

## 4.2 Shadowing Effects Due to Ion-induced Surface Roughness

Past experimental work has shown that ion irradiation at ion incidence angles of ~ 30° – 75° can induce surface roughness through the formation of pits, blisters, cones, ripples, ridges, and facets.[8,13,49–58] The formation of such structure on the micro- and nano-scales is clearly visible in scanning electron microscope images of our post-irradiation Cu target (Fig. 8). The surface roughness has increased from *Ra* ~ 0.1 to ~ 0.5 μm. For these incident angles, theoretical[51] and experimental[13] studies find that a sawtooth-like structure forms on the surface, with one set of facets that are approximately parallel to the irradiating ion beam and another set that are approximately perpendicular (see the discussion and Fig. 1 of Ref. 51 and Fig. 5 of Ref. 13). Our measurement technique does not enable us to study the evolution of the surface of the target as a function of fluence. However, we can compare our study to that of Basu et al.,[13] who report results for 500 eV Ar ions incident on Si at angles of 70° and 72.5°. They find that the ion-generated saw tooth structure converges to a constant large scale structure for a fluence $5 \times 10^{17}$ ions/cm$^2$ and remains approximately constant in shape as the fluence increases to $2 \times 10^{18}$ ions/cm$^2$, the highest value they report. Above $5 \times 10^{17}$ ions/cm$^2$, no further evolution with fluence was seen, such as a flattening of the surface structure. We can scale their results to ours by taking into account two points. First, the sputtering yield at normal incidence for Kr ions on Cu is approximately a factor of ten times larger than that of Ar ions on Si.[59] This implies that we would expect our structure to converge for a factor of ten smaller fluence or $5 \times 10^{16}$ ions/cm$^2$. However, we also need to take into account that the sputtering yield for Kr ions at 45° is approximately half that for 75°.[59] This increases the fluence for convergence to $1 \times 10^{17}$ ions/cm$^2$. Our measurements were performed for a fluence of $1 \times 10^{18}$ ions/cm$^2$. Based on the results of Basu et al.[13], our results are well within the regime for which we expect the surface to have converged to a constant large scale structure.

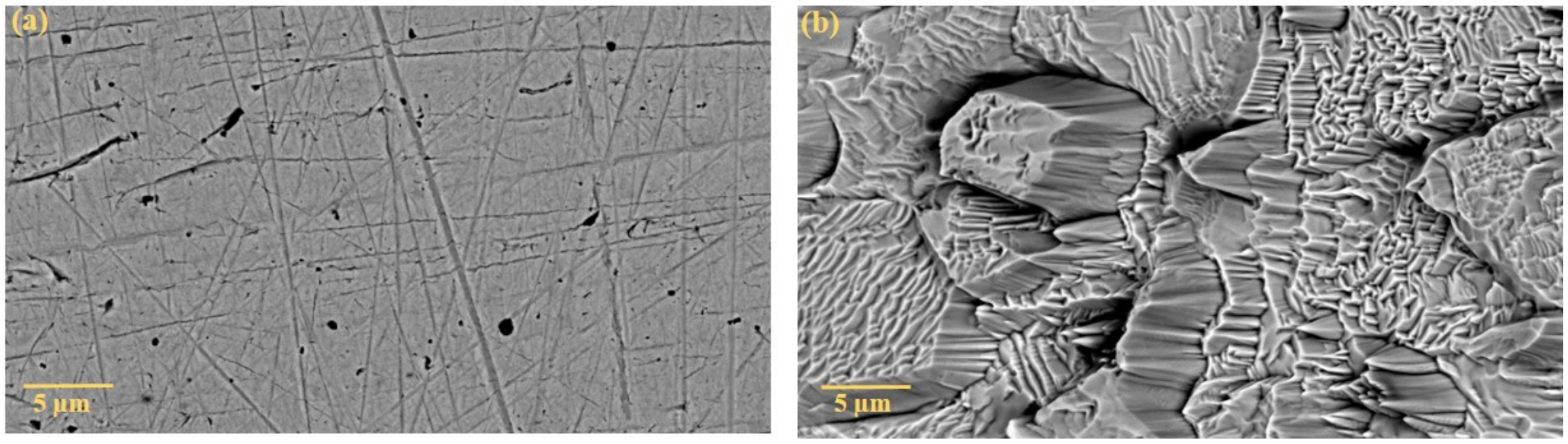

**Fig. 8** Scanning electron microscopy images for our polished Cu target (a) unirradiated and (b) irradiated by $(6.20 \pm 1.24) \times 10^{17}$ ions of 20 keV $Kr^+$. The images were collected using 15 keV primary electrons in the backscattered electron mode at a magnification of 10,000 on a Phenom GLX2. The ion beam velocity vector in (b) is from left to right. The images are from different locations on the sample.

The formation of the sawtooth-like structure leads to shadowing and sputter redeposition that qualitatively explains the azimuthal anisotropy observed in our yield results for $\theta_s = 60°$ and 75°. This is illustrated in Fig. 9. If we assume that the bulk of the sputtering is from the perpendicular facet, then in the forward direction ($\phi_s = 0°$) sputtering is suppressed by shadowing for $\beta = \theta_s = 45° - 90°$ for all points on that facet. In the backward direction ($\phi_s = 180°$) and near the top of the sawtooth, shadowing occurs only for $\alpha = \theta_s \approx 90°$. This increases monotonically to span $\alpha = \theta_s = 45° - 90°$ at the bottom of the sawtooth. Hence for $\theta_s = 60°$ and 75°, sputtering is suppressed in both the forward and backward directions, but more so in the forward direction. Conversely, for $\phi_s$ = ±90° there is no shadowing, leading to the apparent observed sputtering yield enhancement perpendicular to the impacting ions. A quantitative model of surface roughness effects on our measurements is beyond the scope of this work, due to the complexity of the ion-generated surface structure, as can be seen in Fig. 8, but is an important parameter to consider in future projects.

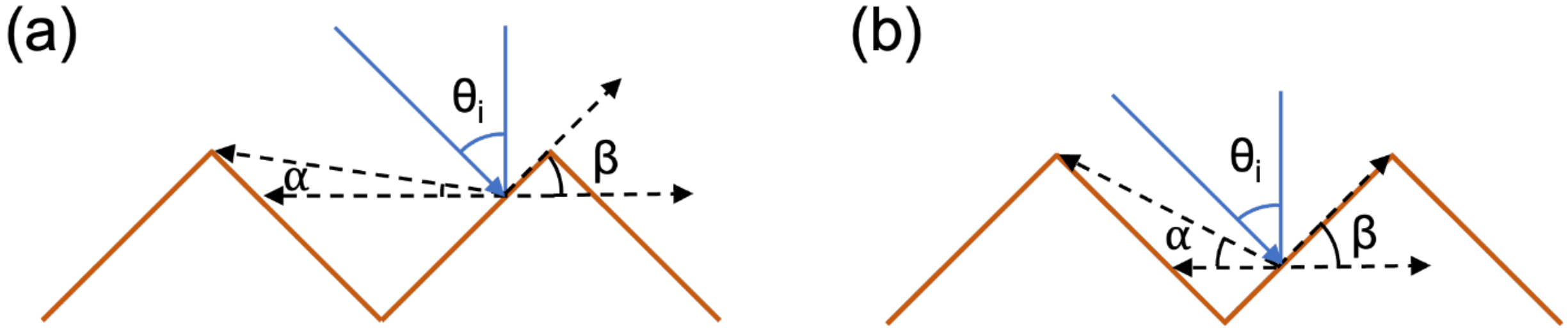

**Fig. 9** Shadowing of the sputtered atoms from a faceted surface for impacts near the top (a) and bottom (b) of the impacted perpendicular facet. The blue arrow indicates the impacting ion velocity vector with an incidence angle of $\theta_i$ = 45° relative to the surface normal (the vertical blue line). The angles β and α show the shadowing of the sputtered atoms in the forward and backward directions, respectively.

## 4.3 Total Sputtering Yield

Despite the significant differences between our experimental and SDTrimSP results for the doubly differential angular sputtering yield, we find surprisingly good agreement in the total sputtering yield. We attribute this agreement to the effects of two competing effects resulting from the ion-induced surface structure – a reduction due to the shadowing and reposition as described above and an enhancement due to the increase of local incidence angle.[49]

In order to determine a total sputtering yield from the experimental results, we have interpolated and extrapolated the measured data points to cover sputtering polar angles of $\theta_s = 0 – 90°$ and azimuthal angles of $\phi_s = 0 – 180°$, both with a step size of 1°. The total experimental yield $Y$ is then calculated by integrating the results over the $\theta_s$-$\phi_s$ plane using:

$$Y = 2\int_{\theta_s=0^\circ}^{90^\circ}\int_{\phi_s=0^\circ}^{180^\circ}(dY/d\Omega)\sin\theta_s \, d\theta_s \, d\phi_s. \qquad \text{Eq. (1)}$$

The interpolated and extrapolated data cover only half of the solid angle above the sample. We assume mirror symmetry in the incident plane and multiply the integrated experimental results by the corresponding factor of 2 seen in Eq. (1).

The experimentally inferred total yield is 23.4 ± 4.7 atoms/ion. SDTrimSP provides a total yield of 21.2 atoms/ion. Experimental total sputtering yields for 20 keV $Kr^+$ on Cu have previously only been carried out for normal incidence ($\theta_i = 0°$), finding a total sputtering yield of ~ 10 atoms/ion,[33] which is in good agreement with SDTrimSP predictions of 10.9 atoms/ion. Our results for $\theta_i = 45°$ are a factor of ~ 2 larger. This increase in the total sputtering yield with increasing $\theta_i$ is comparable to what has been observed experimentally for 1.05 and 45 keV $Kr^+$ on Cu.[60,61]

## 5 Conclusions

We have measured the absolute doubly differential angular sputtering yields for 20 keV $Kr^+$ impacting on polycrystalline Cu at an incidence angle of 45°. Our results show significant differences compared to BCA simulations, demonstrating the potential of doubly differential angular sputtering yields to probe the ion sputtering processes at a fundamental level. In particular, a pronounced azimuthal dependence is observed, which is not predicted by the simulations. Our approach opens up the ability to more fully explore the effect of ion-beam-generated surface roughness in the sputtering process as compared to state-of-the-art experimental methods that report data only for sputtering in the incidence plane. In future work, we will utilize single crystal targets in order to more clearly elucidate sputtering effects resulting from ion-induced surface modifications.

**Supplementary Material**

See the supplementary material for the tabulated measured and simulated absolute doubly differential sputtering yields.

**ACKNOWLEDGEMENTS**

We thank J. E. Lawler and K. A. Miller for stimulating discussions. This work was supported in part by the NASA Solar System Workings grants 80NSCC18K0521 and 80NSCC22K0099. We also acknowledge funding (NIH S10OD016219-01) for the support for setting up the automated weighing system. C. A. Dukes is supported in party by the NSF Division of Astronomical Sciences Astronomy and Astrophysics Grants program under AST-2009365. The authors thank the three anonymous reviewers for their insightful comments and suggestions.

**AUTHOR DECLARATIONS**

**Conflict of Interest**

The authors have no conflicts to disclose.

**Author Contributions**

**Caixia Bu:** Conceptualization (supporting), Data curation (equal); Formal analysis (equal); Funding acquisition (supporting); Investigation (equal); Methodology (equal); Software (equal); Validation (equal); Visualization (equal); Writing – original draft (equal); Writing – review & editing (equal). **Liam S. Morrisse**y: Conceptualization (supporting), Data curation (equal); Formal analysis (equal); Funding acquisition (supporting); Investigation (equal); Methodology (equal); Resources (equal); Software (equal); Validation (equal); Visualization (equal); Writing-original draft (equal); Writing – review & editing (equal). **Benjamin C. Bostick:** Conceptualization (supporting), Data curation (supporting); Formal analysis (supporting); Funding acquisition (supporting); Investigation (supporting); Methodology (supporting); Resources (equal); Supervision (supporting); Validation (supporting); Writing – review & editing (supporting). **Matthew H. Burger:** Conceptualization (supporting); Funding acquisition (supporting); Writing – review & editing (supporting). **Kyle. P. Bowen:** Methodology (supporting); Writing – review & editing (supporting). **Steven. N. Chillrud:** Conceptualization (supporting); Data curation (supporting); Formal analysis (supporting); Funding acquisition (supporting); Investigation (supporting); Methodology (supporting); Resources (equal); Supervision (supporting); Validation (supporting); Writing – review & editing (supporting). **Deborah L. Domingue:** Conceptualization (supporting); Funding acquisition (supporting); Writing – review & editing (supporting). **Catherine A. Dukes:** Data curation (supporting); Formal analysis (supporting); Investigation (equal); Methodology (supporting); Resources (equal); Validation (equal); Visualization (equal); Writing – review& editing (supporting). **Denton S. Ebel:** Conceptualization (supporting); Funding acquisition (supporting); Writing – review & editing (supporting). **George E. Harlow:** Conceptualization (supporting); Funding acquisition (supporting); Writing – review & editing (supporting). **Pierre-Michel Hillenbrand:** Methodology (supporting); Writing – review & editing (supporting). **Dmitry A. Ivanov:** Methodology (supporting); Writing – review & editing (supporting). **Rosemary M. Killen:** Conceptualization (supporting); Funding acquisition (supporting); Writing – review& editing (supporting). **James M. Ross:** Data curation (supporting); Formal analysis (equal); Investigation (equal); Methodology (supporting); Validation (equal); Writing – review & editing (supporting). **Daniel Schury:** Methodology (supporting); Software (equal); Writing – review& editing (supporting). **Orenthal J. Tucker:** Conceptualization (supporting); Funding acquisition (supporting); Writing – review & editing (supporting). **Xavier Urbain:** Methodology (supporting); Writing – review & editing (supporting). **Ruitian Zhang:** Methodology (supporting); Writing – review & editing (supporting). **Daniel. W. Savin:** Conceptualization (lead); Data curation (supporting); Formal analysis (equal); Funding acquisition (lead); Investigation (supporting); Methodology (equal); Project administration (lead); Resources (equal); Software (supporting); Supervision (lead); Validation (equal); Visualization (equal); Writing-original draft (equal); Writing – review & editing (equal).

## DATA AVAILABILITY

The data that support the findings of this study are available within the article.